\documentclass[twocolumn,amsmath,floats,showpacs,nofootinbib]{revtex4}
\usepackage[usenames]{color}
\usepackage{graphicx}
\usepackage{epstopdf}
\usepackage{textcomp}
\usepackage{hyperref}
\usepackage{multirow}
\usepackage{tikz}
\usepackage{rotating}
\hypersetup{colorlinks=true}

\begin{document}

\title{Point-contact spectroscopy of superconducting energy gap in $\rm DyNi_2B_2C$}

\author{I.K. Yanson$^a$, N.L. Bobrov$^a$, C.V. Tomy$^{b,c}$, D.McK. Paul$^b$}
\affiliation{$^a$B.Verkin Institute for Low Temperature Physics and Engineering, 47, Lenin Ave., 310164 Kharkov, Ukraine\\
$^b$Department of Physics, UniIersity of Warwick, CoIentry CV4 7AL, UK\\
$^c$Department of Physics, Indian Institute of Technology, Powai, Mumbai 400 076, India\\
Email address: bobrov@ilt.kharkov.ua}
\published {\href{http://dx.doi.org/10.1016/S0921-4534(00)00227-6}{Physica C} 334 (2000) 33-43}
\date{\today}

\begin{abstract}The superconducting energy gap in $\rm DyNi_2B_2C$ has been investigated using a point-contact technique based on the
Andreev reflection from a normal (N)-superconductor (S) boundary, where N is Ag. The observed differential resistance
$dV/dI$ is well described by the Blonder-Tinkham-Klapwijk (BTK) theory based on the BSC density of states with zero
broadening parameter. Typically, the intensity of the gap structure amounts to several percentage of the normal state
resistance, which is an order of magnitude less than predicted by the theory. For $\rm DyNi_2B_2C$ with $T_c<T_N$ (the Neel
temperature), we found gap values satisfying the ratio of $2\Delta_0/k_BT_c=3.63\pm 0.05$ similar to other superconducting
nickel-borocarbides, both nonmagnetic and magnetic with $T_c\geq T_N$. The superconducting gap nonlinearity is superimposed
on the antiferromagnetic structure in $dV/dI(V)$ which is suppressed at the magnetic field of the order of 3T applied
nominally in the $ab$-plane and temperature $\geq 11~K$. The observed superconducting properties depend on the exact
composition and structure at the surface of the crystal. $\copyright$2000 Published by Elsevier Science B.V. All rights reserved.\\

\end{abstract}

\maketitle
\section{INTRODUCTION}
The rare-earth nickel-borocarbides $\rm RNi_2B_2C$ (R = Y, Lu, Tm, Er, Ho, Dy) attract great attention from researchers in the field of superconductivity because they combine the properties of two conflicting types of long-range order, magnetism and superconductivity, in an easily accessible temperature range. Although their structure is similar to the high-$T_c$ superconductors comprising alternating layers of conducting and (it is believed) superconducting $\rm Ni_2B_2$ planes intercalated with Re-C layers, their electronic properties are surprisingly isotropic due to the strong carbon links along the tetragonal $c$-axis. Moving along the row of increasing de Gennes factor $(g - 1)^2J(J + 1)$, starting from nonmagnetic Y- and Lu-compounds, their Neel temperatures $T_N$ increase almost linearly while their superconducting transition temperatures $T_c$ decrease from 16.5 down to 6~$K$. Hence, the conduction electrons interact strongly with the f-electrons of the rare-earth ions. Recently, Cho et al. \cite{1} have reported the complete breakdown of de Gennes scaling in $\rm Ho_{1-x}Dy_xNi_2B_2C$ and $\rm Lu_{1-x}Dy_xNi_2B_2C$ pseudoquaternaries which could be explained by the important role of magnons in these
compounds. They suggest that the magnon excitations are responsible for suppressing the superconducting transition in these pseudoquaternaries. Another closely related magnetic superconductor, the Ho-compound, which has the same ground state crystal and magnetic structure, similar mass of the R ion and ordered magnetic moments, was studied by means of point-contact spectroscopy in Refs. \cite{2,3,4}. It was shown that $\rm HoNi_2B_2C$ possesses two distinct superconducting states connected with the changes in magnetic order \cite{5,6}. At $T_c\leq8.5\ $K, the spiral magnetic order appearing along with superconductivity leads to the non-BCS behavior in the density of quasiparticles states (DOS), while below $T_N\simeq 6\ K$, the simple commensurate antiferromagnetic order results in a BCS-DOS and a BCS-temperature-dependence with an effective critical temperature $T_c^*\simeq T_N$ \cite{2}. The reason for this behaviour is that at $T\leq T_c^*$, the effective internal magnetic field along superconducting $\rm Ni_2B_2$ planes cancels \cite{5}. The PCS-study of the electron-boson-interaction spectral function of $\rm HoNi_2B_2C$ reveals low-frequency phonon, magnon\footnote{Magnon excitaton peaks in PCS spectra become observable at the magnetic transition temperature and strongly grow in intensity with further decrease of the temperature.} and crystal electric field excitation (CEF) branches \cite{1,7,8} which are greatly overlapped and might play an important role on the mechanism for superconductivity.

The superconducting energy gaps have been measured by tunneling in two nonmagnetic compounds $\rm Lu/YNi_2B_2C$ \cite{9,10} which give $\Delta_0\simeq 2.2\ meV$ with $2\Delta_0/k_BT_c\simeq 3.2-3.5$. Unfortunately, the tunneling characteristic was not of an excellent quality which forces one to suspect that at the surface, the superconductivity is partially depressed. The point-contact measurements of superconducting energy gaps in Y- \cite{11,4,12}, Ho- \cite{2} and Er-compounds \cite{12} give $2\Delta_0/k_BT_c = 3.63\pm 0.05$ if for Ho-, one takes the "lower" $T_c^* = 6\ K$. To our knowledge, there are no tunneling measurements of the superconducting energy gap in any magnetic nickel-borocarbides. It should be noticed that there is no structure in the tunneling spectrum above the gap voltage in $\rm Lu/YNi_2B_2C$, which would allow one to infer the Cooper-pair-mediation mechanism.

In this work, we present measurements of the superconducting energy gap in the magnetic superconductor $\rm DyNi_2B_2C$ \cite{13,14}. As is well known, PCS spectra of quasiparticle DOS utilizes the Andreev reflection phenomena at the normal (N)-superconductor (S) boundary inside the N-c-S (c-constriction) point contact. The superconducting transition of $\rm DyNi_2B_2C$ is at $T_c = 6\ K$, deep inside the antiferromagnetic state which occurs at $T_N = 10.5\ K$. We show that the $2\Delta_0/k_BT_c$ ratio for $\rm DyNi_2B_2C$ has the value of 3.6 stated above for other superconducting
$\rm RNi_2B_2C$ \cite{12}, and the DOS and temperature dependence obeys the BCS law. The upper critical field amounts to approximately 0.9~$T$ for a field aligned nominally within the $ab$-plane. The superconducting gap nonlinearity is superimposed on the AFM structure in the current-voltage characteristics. The magnetic structure is suppressed at a field of the order of 3~$T$. The observed superconducting properties are influenced by the composition and crystal structure at the surface of the crystal and can be easily destroyed by slight disorder inside a contact.  $\rm DyNi_2B_2C$ is the most difficult material to make superconducting, since the superconductivity appears to critically depend on the exact C content. This is true for polycrystalline and single crystal samples. The description of PCS study of phonon, CEF, and magnon low-frequency spectral bands at the over-gap energies will be reported in a following article \cite{15}.

\section{Methodical details}

\begin{figure}[]
\includegraphics[width=6cm,angle=0]{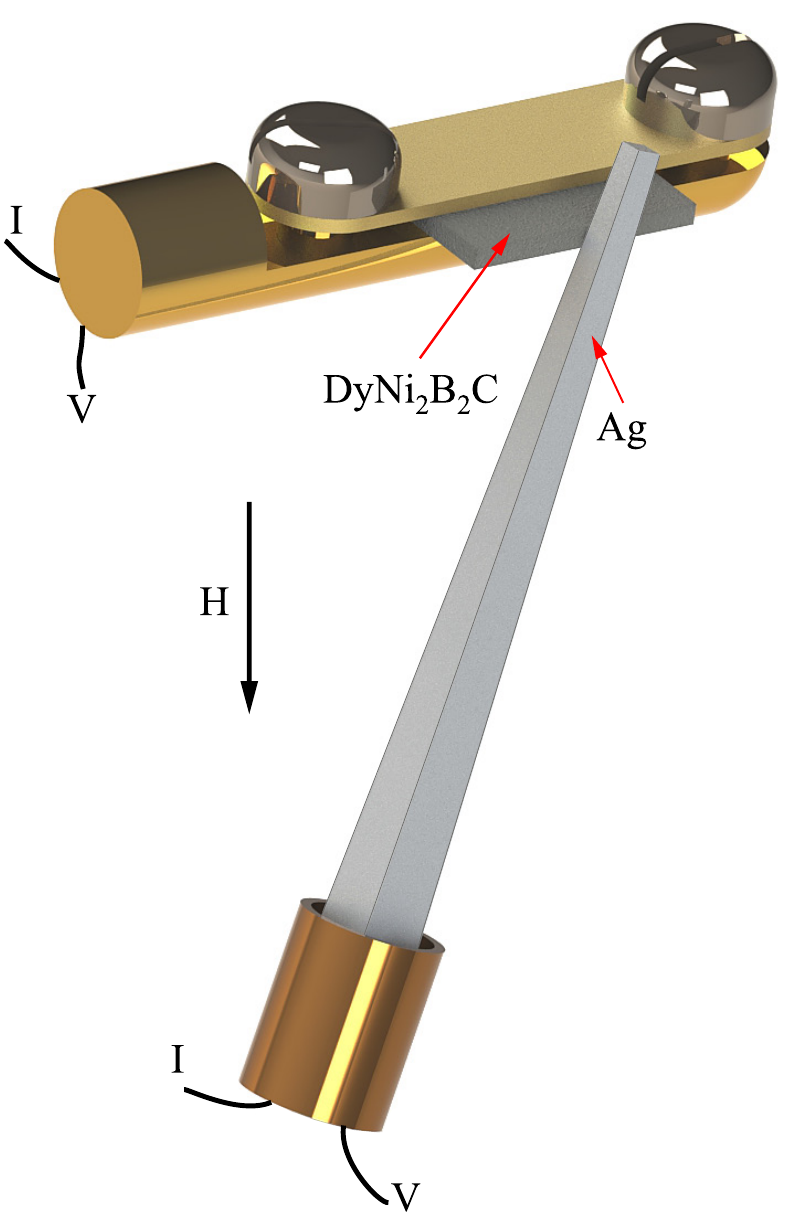}
\caption[]{Experimental setup for measuring single crystal of Dy-compound.}
\label{Fig1}
\end{figure}
The schematic setup for sample measurements is shown in Fig.\ref{Fig1}. A small single crystal of $\rm DyNi_2B_2C$ is put into electrical contact with the sharp edge of an Ag rod using a mechanism operated from outside the cryostat. Many spots of the crystal can be probed as well as many points on the Ag rod. The orientation of the crystal was such that the current flows nominally along the $ab$-plane. The magnetic field, if necessary, is applied perpendicular to the current and nominally parallel to the $ab$-plane. Unfortunately, the precise angle between $ab$-plane and H and within the $ab$-plane cannot be determined in this geometry. The current-voltage characteristic along with the first and second derivatives are measured by means of a lock-in technique.

\begin{figure}[]
\includegraphics[width=8.5cm,angle=0]{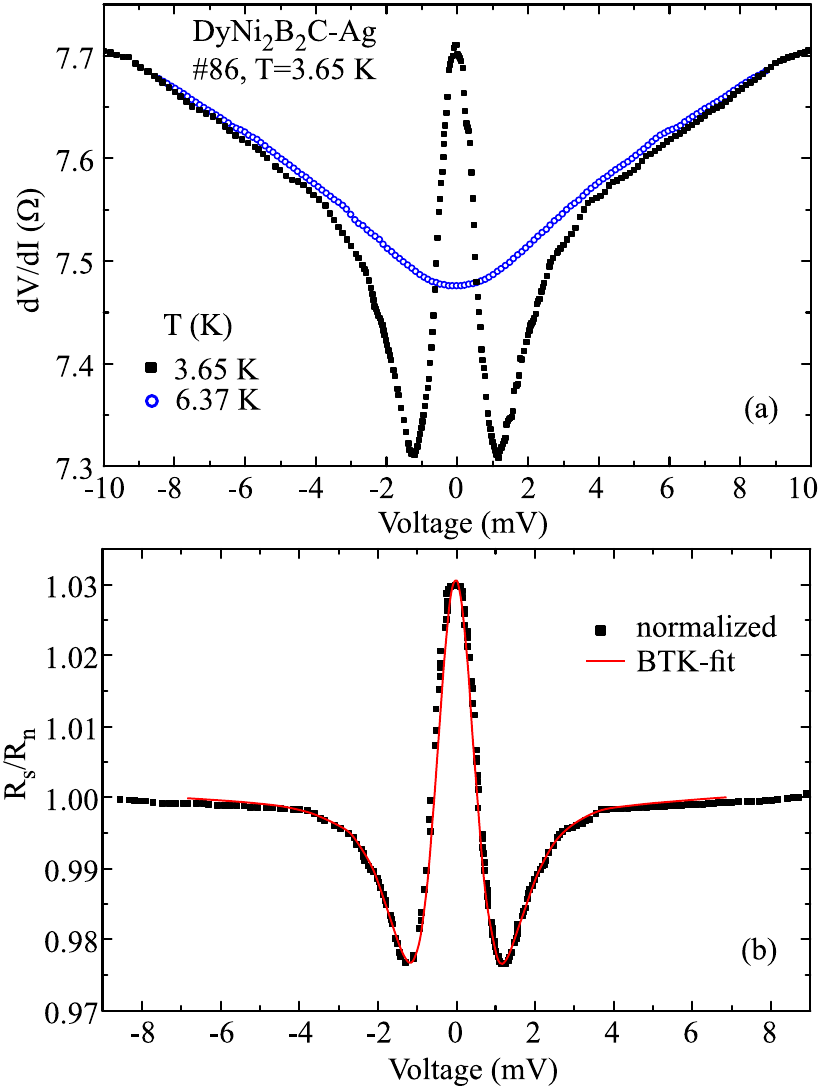}
\caption[]{(a) Symmetrized $dV/dI$ spectra of $\rm DyNi_2B_2C-Ag$ contact in the superconducting ($T = 3.65\ K$) and normal state ($T = 6.37\ K$). (b) Normalized experimental data points with the BTK fit (solid curve).}
\label{Fig2}
\end{figure}

\begin{figure}[]
\includegraphics[width=8.5cm,angle=0]{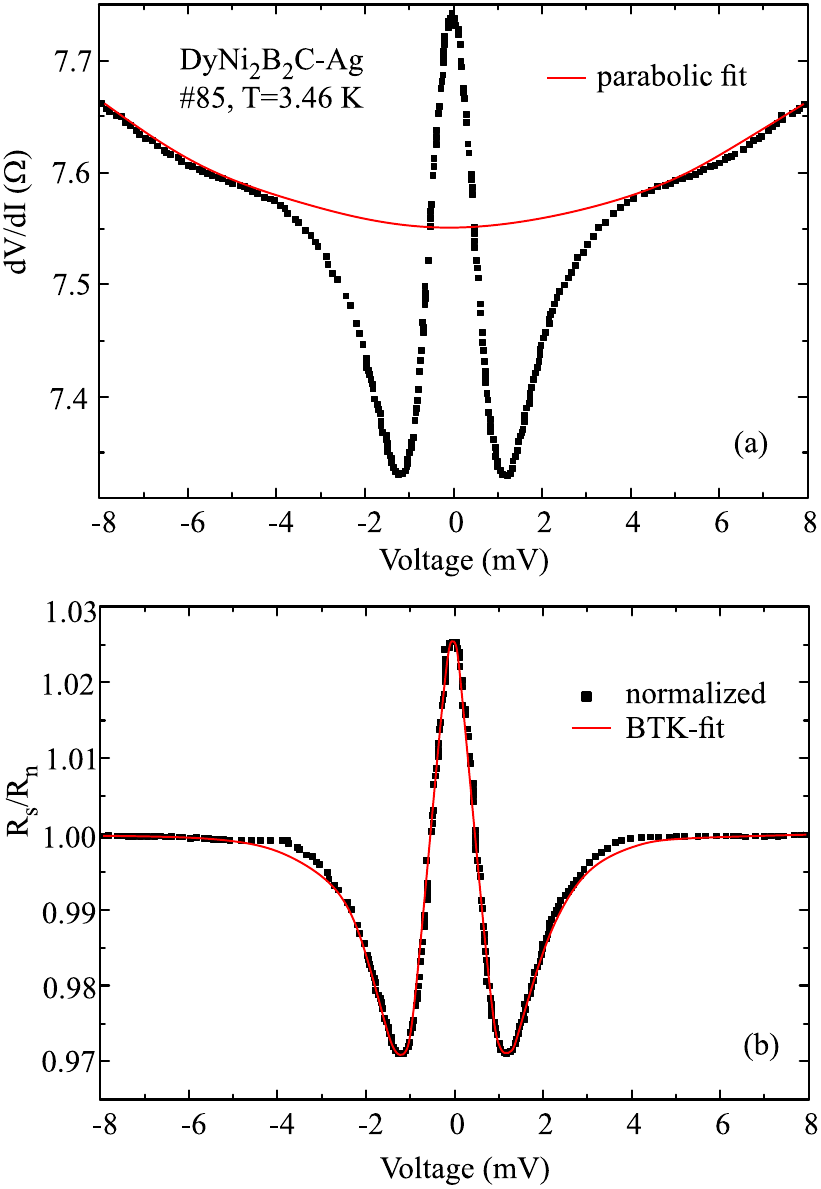}
\caption[]{The same as in Fig.\ref{Fig2} but for the neighboring contact at slightly lower temperature. Now, as a background, the parabolic fit is used (see also Fig.\ref{Fig4}).}
\label{Fig3}
\end{figure}

In Figs.\ref{Fig2} and \ref{Fig3}, we show how we subtract the normal-state background and normalize the $R(V) = dV/d I(V)$ spectra. Firstly, the spectra are symmetrized through the formula $dV/dI_{sym} = (1/2) [dV/dI(V) + dV/dI(-V)]$. The reason is that we fit them by the Blonder-Tinkham-Klapwijk (BTK) theory which is symmetric with respect to $\pm V$. In many cases, the asymmetry of experimental curve is negligible, which is illustrated by the as-measured data in Figs.\ref{Fig7},\ref{Fig8}. Then, two methods for the subtraction of a normal-state background may be used. Fig.\ref{Fig2}(a) illustrates the case when we have measured $R(V)$ at the onset superconducting temperature $T^*\geq T_c$ and its asymptotic behavior at high voltages coincides with that for the low-temperature curve. Usually, this "native" normal-state background has a parabolic shape near $V=0$. Fig.\ref{Fig2}(b) displays the
result of normalizing $dV/dI$ in the superconducting state and fitting it with the BTK theory \cite{16} based on the Andreev reflection phenomena \cite{17}. It is interesting to note that throughout all this work, we held the $\Gamma$-parameter, introduced by Dynes et al. \cite{18}, equal to zero, since finite $\Gamma$ did not improve the fit. For some samples (not discussed in this paper), a small nonzero $\Gamma$ can improve the fit. In any case, this does not change the gap value within the limits of error bar (see Fig.\ref{Fig13}).

Another case is shown in Fig.\ref{Fig3}, which corresponds to either the absence of the normal-state
curve (due to, for instance, contact instability) or a change in voltage behavior which modifies the low-temperature curve at high voltages. In this case, we usually approximate the normal-state background by parabola matching the high-voltage branches of the curve in question (Fig.\ref{Fig3}(a)). Again, the result of such normalization is given in Fig.\ref{Fig3}(b) with the corresponding BTK-fit. From the fit, we determine the barrier parameter $Z$ \cite{16} and the energy gap at the given temperature $\Delta (T)$.

\begin{figure}[]
\includegraphics[width=8.5cm,angle=0]{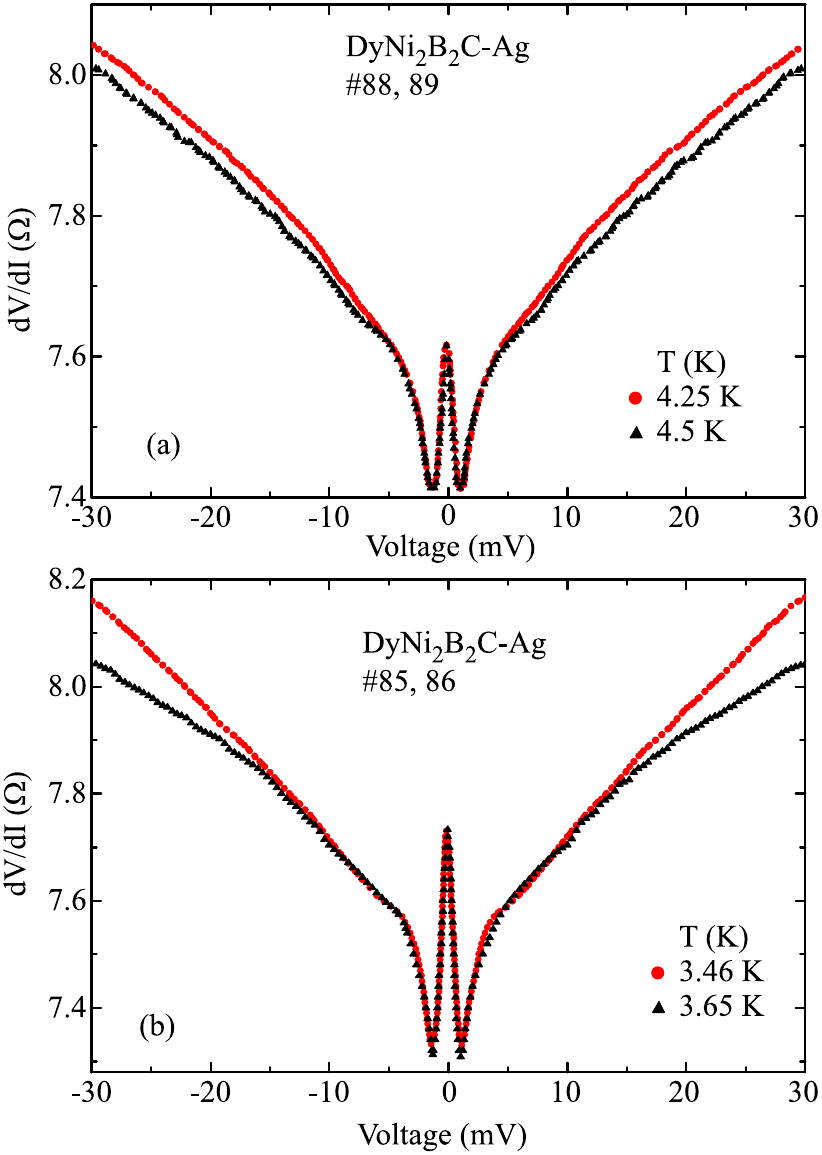}
\caption[]{A pair of $dV/dI$ spectra at similar temperatures showing a considerable difference in the background.}
\label{Fig4}
\end{figure}

\begin{figure}[]
\includegraphics[width=8.5cm,angle=0]{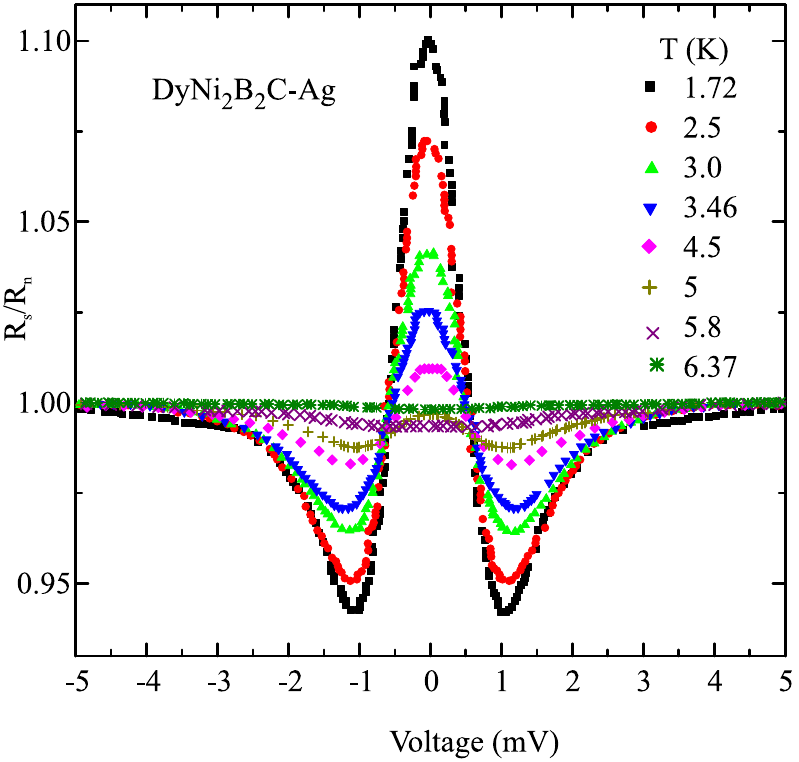}
\caption[]{Temperature dependence of normalized $dV/dI$ spectra for the first representative series of $\rm DyNi_2B_2C-Ag$ symmetrized and normalized.}
\label{Fig5}
\end{figure}

\begin{figure}[]
\includegraphics[width=8.5cm,angle=0]{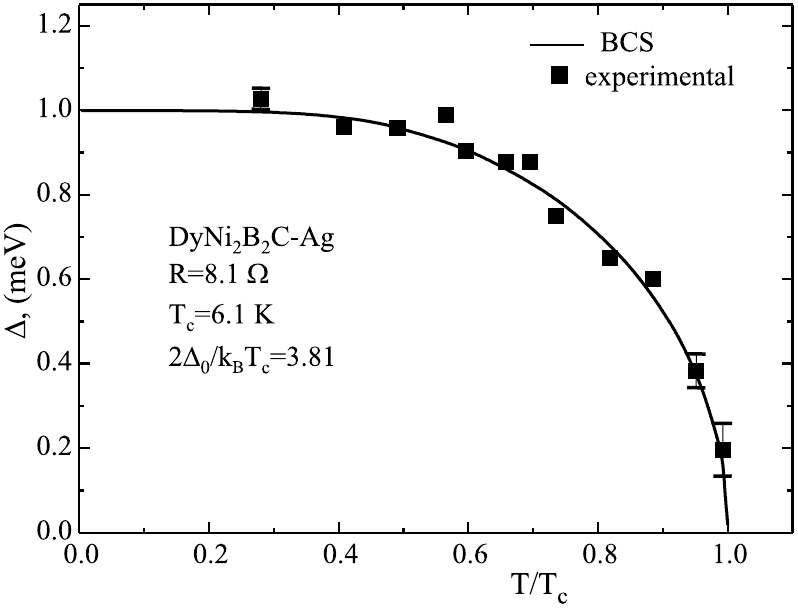}
\caption[]{Temperature dependence of the superconducting energy gap determined by BTK-fit to the first series of $\rm DyNi_2B_2C-Ag$ contacts.}
\label{Fig6}
\end{figure}

Below, we present detailed results for two typical $\rm DyNi_2B_2C-Ag$ contacts and some other situations, using which, we can discuss some additional methodical details.

\section{Experimental results}

\begin{figure}[]
\includegraphics[width=8.5cm,angle=0]{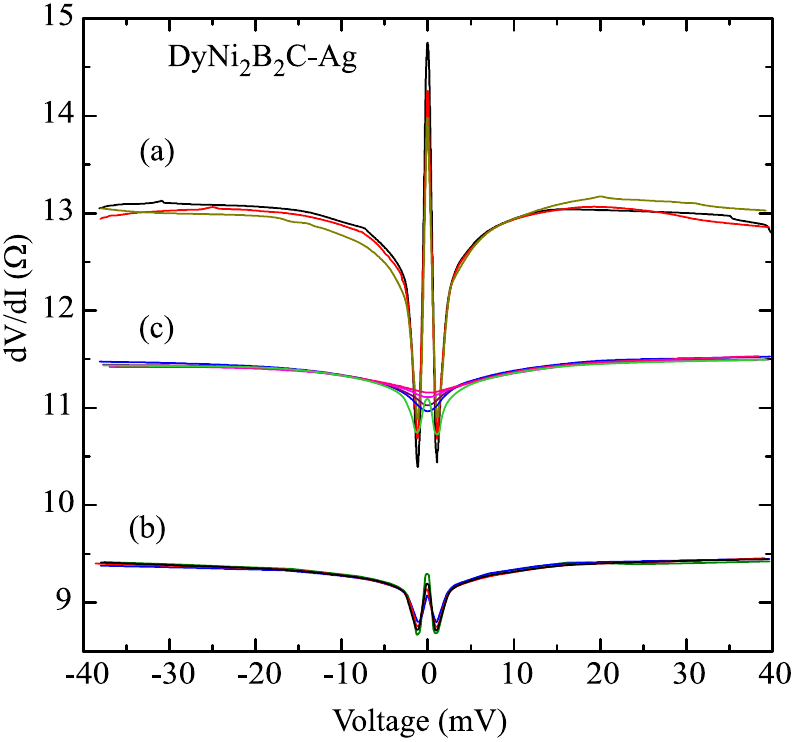}
\caption[]{Temperature dependence of the $dV/dI$ spectra for the second representative series of the $\rm DyNi_2B_2C-Ag$ contacts as measured. The groups of curves (a), (b) and (c) are recorded in the temperature range 1.7-2.61, 2.83-4.26 and 4.52-9.0~$K$, respectively.}
\label{Fig7}
\end{figure}

Most contacts reveal featureless spectra showing neither superconducting gap structure nor any other nonlinearities. We suspect that the metal under the contact has been severely spoiled in the process of making the contact. Also, we suspect that the surface of the crystal degrades with time when maintained at ambient conditions. To increase the possibility of obtaining good contacts, we break the crystal and try to measure it as quickly as possible. That the observed characteristics are very sensitive of minor disturbances is demonstrated in Fig.\ref{Fig4}(a) and \ref{Fig4}(b) where a pair of symmetrized spectra at similar temperatures is shown. These small changes of temperatures lead to noticeable changes in the background at high voltages. The superconducting energy gaps at low voltages determined by BTK-fitting are different as well (see below). These curves illustrate that the normal-state background should be chosen differently for each curve, even when they are similar in
temperature. Representative normalized curves for the same series of measurements are plotted in Fig.\ref{Fig5}. These curves, together with others not shown in the graph for clarity, were BTK-fitted to yield the energy gap as a function of temperature (Fig.\ref{Fig6}). Note that the typical deviation from ohmic behavior amounts to only a few percentage, while for the nonmagnetic compound (see, for example,  $\rm YNi_2B_2C-Ag$ in Ref. \cite{11}), the Andreev-reflection increase of conductance at zero bias equals a few tens of percents, as it should be for a conventional S-c-N contact according to the theory. To within the scatter in the data, the temperature dependence of the superconducting energy gap follows the law predicted by BCS (solid curve in Fig.\ref{Fig6}). Take note here (Fig.\ref{Fig4}) of the two data points at $T/T_c$=[0.567, 0.598] and [0.697, 0.738], which show noticeable different energy gaps reflecting changes in the material properties under the contact. For this particular contact, $T_c=6.1\ K$ with $2\Delta_0/k_BT_c=3.8$ (demonstrating small variations in $T_c$ and $\Delta_0$ from place to place on the sample surface).
\begin{figure}[]
\includegraphics[width=8.5cm,angle=0]{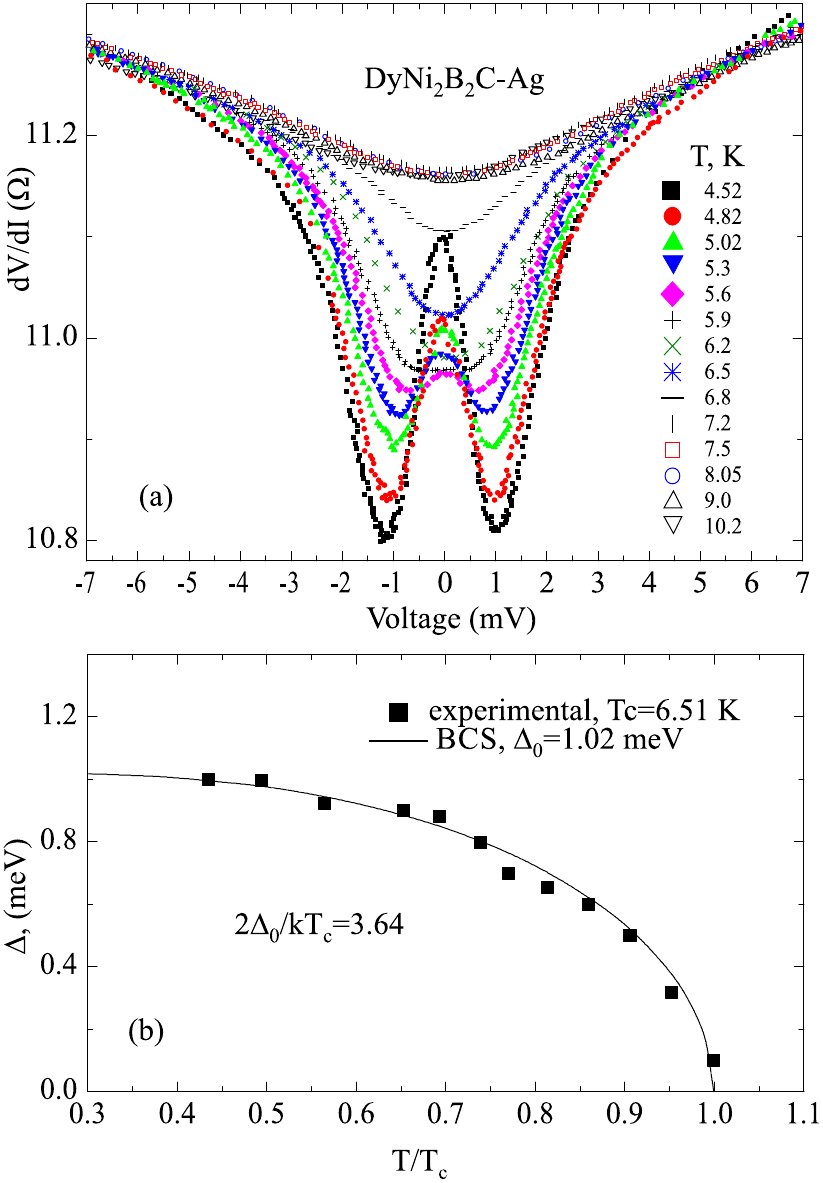}
\caption[]{(a) As-measured $dV/dI$ spectra of the second series for $\rm DyNi_2B_2C-Ag$ contacts in the temperature range 4.52-10.2~$K$. (b) Temperature dependence of the superconducting energy gap for groups of curves (b) and (c) from Fig.\ref{Fig7}.}
\label{Fig8}
\end{figure}

\begin{figure}[]
\includegraphics[width=8.5cm,angle=0]{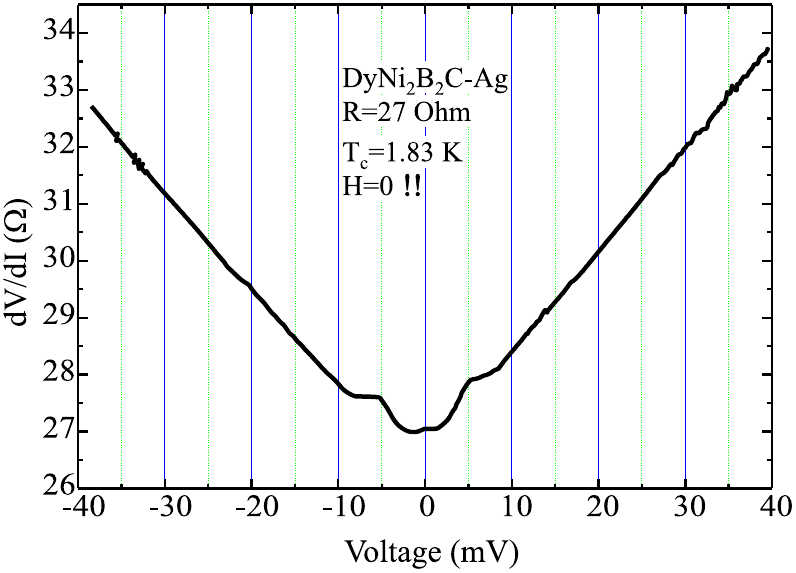}
\caption[]{Antiferromagnetic singularity near zero bias for the $\rm DyNi_2B_2C-Ag$ contact with no traces of superconductivity.}
\label{Fig9}
\end{figure}

\begin{figure}[]
\includegraphics[width=8.5cm,angle=0]{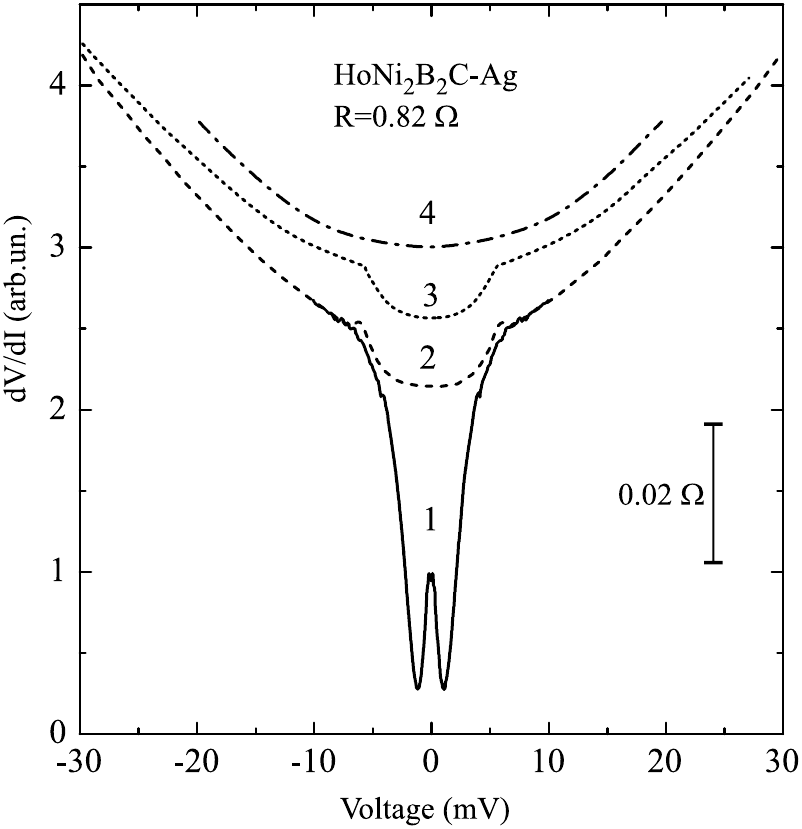}
\caption[]{The $dV/dI$ spectra for $\rm HoNi_2B_2C-Ag$ contact in superconducting state (curve 1: $T = 1.6\ K$, $H=0$), in normal but antiferromagnetic state (curves 2 and 3: $T=1.6\ K$, $H=0.5\ T$; and $T = 4.2\ K$, $H = 0.7\ T$, respectively), and normal and paramagnetic state (curve 4: $T = 10\ K$, $H = 0$).}
\label{Fig10}
\end{figure}

\begin{figure}[]
\includegraphics[width=8.5cm,angle=0]{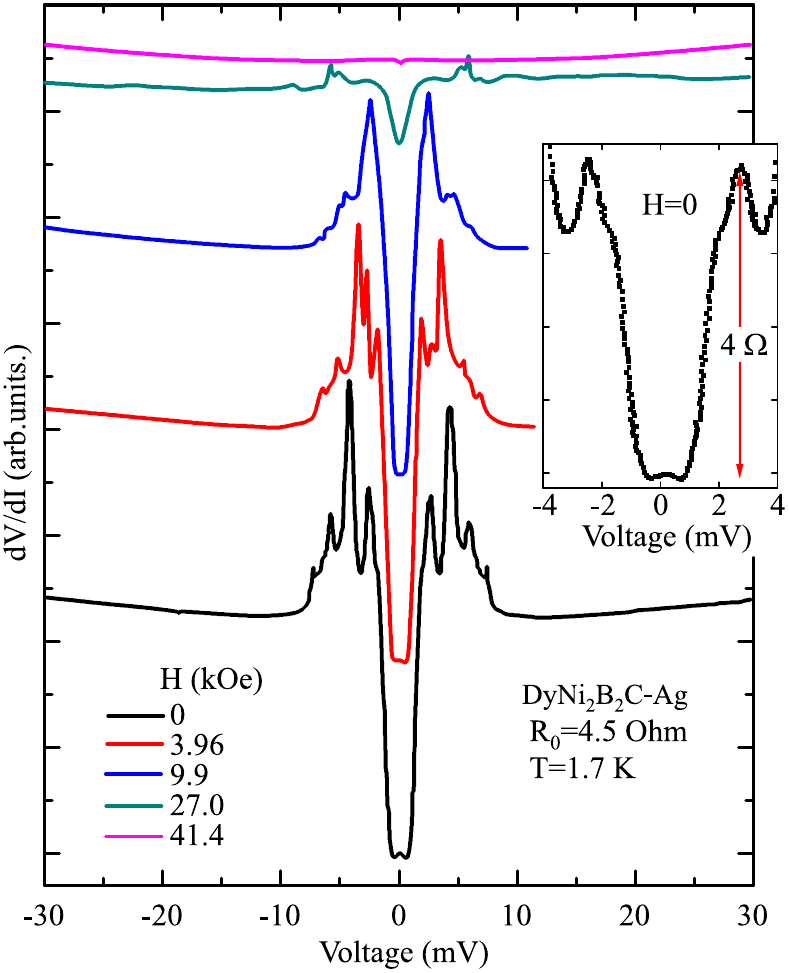}
\caption[]{The $dV/dI$ spectrum with an anomalous increase in the intensity of the conductance due to superpositioning of the superconducting and antiferromagnetic structures near zero bias. In the inset, the enlarged portion of the curve for low biases is shown.}
\label{Fig11}
\end{figure}

\begin{figure}[]
\includegraphics[width=8.5cm,angle=0]{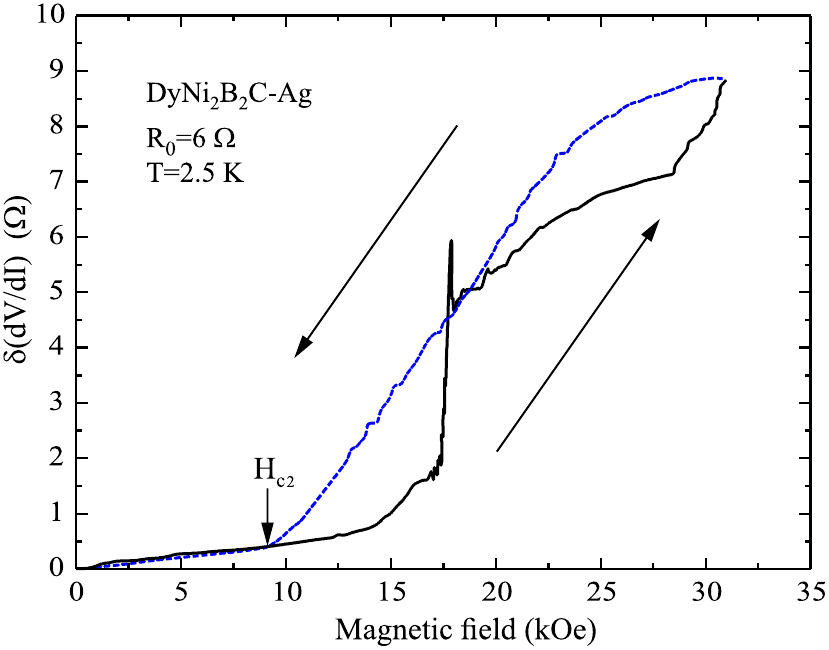}
\caption[]{Dependence of the change of zero-bias resistance for $\rm DyNi_2B_2C-Ag$ contact from Fig.\ref{Fig11} under the influence of a magnetic field. The arrows in opposite directions show the sweep direction for the magnetic field.}
\label{Fig12}
\end{figure}

\begin{figure}[]
\includegraphics[width=8.5cm,angle=0]{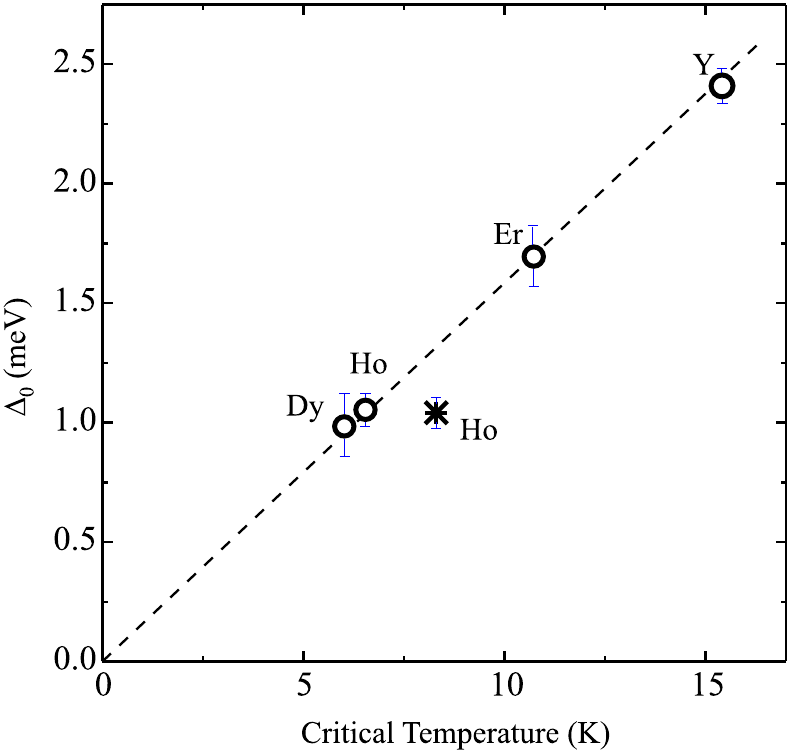}
\caption[]{Zero-temperature energy gap as a function of critical temperature for several nickel-borocarbides. For Ho-compound, two critical temperatures are shown: the lower corresponds to the commensurate antiferromagnetic transition.}
\label{Fig13}
\end{figure}

One of the highest contact conductance conditions (about 20\%) at $eV\simeq\Delta$ at low temperatures is shown in Fig.\ref{Fig7}(a). However, raising the temperature above 2.61~$K$ leads to a step-like decrease in the normal-state resistance from 13 to 9.5~$\Omega$ (curve (b)) with a drastic decrease in the observed nonlinearity down to 6\%. A further increase of temperature (curve (c)) again leads to a step-like change of the normal resistance (up to 11.5~$\Omega$) with the same magnitude of the nonlinearity (of the order of a few \%) taking into account that with increasing the temperature, the increase of conductance at $eV\simeq\Delta$ monotonously diminishes. Fitting these spectra with the BTK procedure shows that for curve (a), $\Delta= 1.08-1.15\ meV$ while for curves (b) and (c), $\Delta(T)$ is well approximated with BCS temperature dependence with $T_c=6.51\ K$ and $\Delta_0=1.02\ meV$. The latter gives a $2\Delta_0/k_BT_c$ ratio equal to 3.64 and a typical DOS nonlinearity of several percentages. Fig.\ref{Fig8}(a) presents the curves of Fig.\ref{Fig7}(c) as measured, while Fig.\ref{Fig8}(b) displays $\Delta(T)$-dependence for series (b) and (c).

All these observations show that the observed spectra are critically dependent on the state of the material inside a contact. Beside the composition and crystal structure order of compound at the surface, small changes of contact pressure or/and orientation of contact axis with respect to the crystallographic axis may also play a role in observing a variety of $dV/dI$ spectra.

We close this section by mentioning two anomalous spectra. The first (Fig.\ref{Fig9}) shows no superconducting energy gap though some nonlinearity at low biases is quite visible. This structure is evidently due to the antiferromagnetic order (AFM) because it disappears at temperature above 11~$K$ and for magnetic fields above 3~$T$. It seems that the magnetic order is detrimental to superconductivity. Some traces of this structure can be noticed in Fig.\ref{Fig4}(a) and (b) and for many other contacts. We noticed that the lower the biases where AFM structure is observed (in other words, the softer the magnon spectrum), the more the superconducting structure is suppressed. This property coincides with the conclusion inferred in Ref.\cite{1}.

Remarkably, the same magnetic structure is observed in the AFM state of Ho-compound. In Fig.\ref{Fig10}, we show how the superconducting energy gap is transformed to AFM wide minimum near zero bias when the requisite magnetic field is applied to destroy the superconductivity (curve 1 $\rightarrow$ curve 2). This AFM structure is preserved for higher temperature and field (curve 3) and disappears only above the magnetic transition temperature ($T_m = 8-9\ K$ for spiral magnetic structure for $\rm HoNi_2B_2C$ at zero magnetic field) or above the magnetic saturation field (2-3~$T$ at low temperatures). Unfortunately, for the Ho-compound, the superconducting transition temperature $T_c$ and the magnetic transition temperature $T_m$ almost coincide which hinders their separation when constructing the phase diagram by means of PCS (such difficulty is seen in Ref.\cite{2}).

The second anomalous spectrum (Fig.\ref{Fig11}) shows a giant increase of conductivity near zero bias (about 100\%) which disappears only at about $H=3-4\ T$. This is the saturation magnetic field for $\rm DyNi_2B_2C$ in between the axis [110] and [100], which is noticeably higher than $H_{c2}\simeq 0.9\ T$. It is interesting that the zero-bias point-contact resistance shows hysteresis while reversing the direction of field change (Fig.\ref{Fig12}) \cite{19}, which points to the first order transition at about $2T$ at $T=2.5\ K$. We interpret this anomalous spectrum as the rare case where the large superconducting and magnetic nonlinearities coincide, yielding such a giant increase of zero-bias conductance.

\section{Discussion}

From the quasiparticle DOS structure and its temperature dependence, $\rm DyNi_2B_2C$ behaves like a BCS superconductor with a moderately strong electron- boson (phonon?) interaction. It fits well with the other $\rm RNi_2B_2C$ superconductors with $2\Delta_0/k_BT_c =3.63\pm 0.05$ (Fig.\ref{Fig13}; see Ref. \cite{12}). This is quite surprising, taking into account that among these compounds, there is a nonmagnetic one (Y-compound) and crystals with quite different magnetic order (Er- \cite{20,21}, Ho- \cite{5,6}, Dy-compounds \cite{22}). The necessity of taking "low" $T_c^* = T_N \simeq 6.5\ K$ for Ho- for this $\Delta_0(T_c)$-law is proved by the fact that all the other magnetic superconductors have the magnetic structure, yielding antiferromagetic compensation on the Ni planes, while the misalignment of the adjacent ferromagnetic planes in $\rm HoNi_2B_2C$ at $6.5\leq T\leq 8.5\ K$ destroys the antiferromagnetic compensation on the $\rm Ni_2B_2$ layers. Note that although the marked dependence of Neel and superconducting transition temperatures on de Gennes factor suggests a strong interaction between the conduction electrons and the rare-earth magnetic moments, the DOS in the superconducting state is well fitted by a zero $\Gamma$-parameter in the AFM phase over the whole temperature range. This means that the depairing interactions are not visible in the superconducting phase. This is probably because there are no disordered magnetic moments in the AFM superconducting phase.

The magnitude of the low-temperature conductance increase near zero-bias in typical cases amounts only to a few percentage of the normal conductance which is an order of magnitude less than predicted by the BTK theory. On the other hand, as we have
mentioned above, a conventional increase in conductance is observed for point-contacts with nonmagnetic $\rm YNi_2B_2C$ compound. We speculate that this may be due to the peculiar mechanism of Andreev reflection from the magnetic superconductor, which is not theoretically treated yet. It is known that the intensity of the Andreev reflection is diminished at the boundary between a conventional superconductor and a ferromagnetic metal depending on the polarization of the spin in the ferromagnet \cite{23,24}. Although the internal magnetic field cancels to zero on the length scale of the superconducting coherence length ($\xi_0\simeq 70\ \text{\AA}$) in the antiferromagnet, some electron polarization may remain depending upon the angle between the direction of the electron momentum and the $ab$-plane when the magnetic moments of the rare-earth ions are ordered ferromagnetically in each particular plane. This may be the cause of the observed suppression of Andreev reflections, as well as the variety of particular contact characteristics. Interestingly, in $\rm ErNi_2B_2C$, where the magnetic moments order antiferromagnetically along the $ab$-plane, the preliminary results point to a conventional intensity for the Andreev reflection structure \cite{12}. Surprisingly, there is no pronounced anisotropy of $H_{c2}$ along and perpendicular to the $c$-axis, like for example, in $\rm URu_2Si_2$, which may be because only the electrons with moments inside the small $(\delta_c)/\xi_0$ ($(\delta_c)$-elementary cell distance along $c$-axis) angle are magnetically polarized.
\section{Acknowledgements}
I.K.Y. is grateful for the financial support from the International Science Foundation (Soros Foundation).

\end{document}